\documentclass[preprint]{aastex701}

\usepackage{multirow}

\usepackage{soul}
\usepackage{color}
\usepackage{cancel}

\begin{document}

\title{Effects of the Background Magnetic Field on Flux Rope Eruptions}

\author[0009-0009-2176-6017]{Xianyu Liu} 
\affil{Department of Climate and Space Sciences and Engineering, University of Michigan, Ann Arbor, MI 48109, USA}
\email{xianyu@umich.edu}

\author[0000-0003-0176-4312]{Spiro K. Antiochos} 
\affil{Department of Climate and Space Sciences and Engineering, University of Michigan, Ann Arbor, MI 48109, USA}
\email{spiro.antiochos@gmail.com}

\author[0000-0002-6118-0469]{Igor V. Sokolov} 
\affil{Department of Climate and Space Sciences and Engineering, University of Michigan, Ann Arbor, MI 48109, USA}
\email{igorsok@umich.edu}

\author[0000-0001-9360-4951]{Tamas I. Gombosi}
\affil{Department of Climate and Space Sciences and Engineering, University of Michigan, Ann Arbor, MI 48109, USA}
\email{tamas@umich.edu}

\author[0000-0001-5260-3944]{Bart van der Holst} 
\affil{Astronomy Department, Boston University, Boston, MA 02215, USA}
\email{bartvand@bu.edu}

\author[0000-0001-9114-6133]{G\'abor T\'oth}
\affil{Department of Climate and Space Sciences and Engineering, University of Michigan, Ann Arbor, MI 48109, USA}
\email{gtoth@umich.edu}

\author[0000-0001-9114-6133]{Nishtha Sachdeva}
\affil{Department of Climate and Space Sciences and Engineering, University of Michigan, Ann Arbor, MI 48109, USA}
\email{nishthas@umich.edu}

\author[0000-0003-3936-5288]{Lulu Zhao}
\affil{Department of Climate and Space Sciences and Engineering, University of Michigan, Ann Arbor, MI 48109, USA}
\email{zhlulu@umich.edu}

\begin{abstract}

Solar eruptive events are generally believed to involve magnetic flux ropes (MFR), formed either in the pre-eruptive phase of the event or during the eruption itself. These MFR eruptions exhibit significant complexity and variations due to the interplay of the physical mechanisms involved, in particular magnetic reconnection and ideal instabilities. This work considers the effect of the background magnetic field on the nature of eruptions with pre-existing MFRs. We used a new MHD model to simulate the whole MFR eruption process, including the pre-eruptive stage and the initiation. Three simulations were performed, all of which used an identical bipolar active region, but with different background magnetic fields in the three cases. The simulations resulted in two successful eruptions (CMEs) and one failed eruption (a confined flare). We analyzed the energetics and the acceleration of the MFR in detail, and found a transition to a rapid exponential rise phase in two of the simulations. We also calculated the criterion for the torus instability and the timing of the breakout and flare reconnections. Our results show that the rapid exponential rise phase is likely due to breakout reconnection. We conclude that a background field antiparallel to the active-region field lowers the magnetic free-energy threshold for eruption; but, does not guarantee a successful eruption. We also found that an antiparallel background field leads to faster flare reconnection, but of shorter duration. Our findings underscore the importance of the background magnetic field in understanding CMEs.

\end{abstract}

\section{Introduction}\label{section_intro}

In recent years, there has been a growing effort towards modeling coronal mass ejection (CME) events that are well matched to observations \citep[e.g., ][]{Torok2018,Downs2021,Fan2024}. However, as noted by \citet{Torok2023}, despite extensive work on CMEs, our understanding of the fundamental physical processes of CMEs is still lacking. An important aspect is that intense debate remains over the mechanism for CME initiation. 

CMEs are widely recognized as the most spectacular manifestations of solar magnetic activity \citep[e.g., ][]{Webb2012}. CMEs eject tremendous amounts of magnetic flux and mass from the solar corona at speeds ranging from hundreds to thousands of km/s \citep[][]{Yashiro2004}. Since the first direct observation of CMEs, numerous efforts have been made to understand their physical nature and predict their space weather effects \citep[see, ][]{Chen2011,Webb2012,Manchester2017}. Most CMEs originate from filament channels (FCs) that are located above a polarity inversion line (PIL) of the photospheric magnetic field \citep[][]{McIntosh1972,Martin1998}. The pre-eruptive magnetic configuration can be either a magnetic flux rope \citep[MFR, e.g.,][]{TD1999} or a sheared magnetic arcade \citep[SMA, e.g.,][]{Antiochos1994}. Observations show that the pre-eruptive configuration can stay in equilibrium for a long duration up to several months \citep[e.g.,][]{Gibson2006}. The eruption is triggered when the evolution of the configuration disrupts such an equilibrium \citep[e.g.,][]{Forbes2000}. After the eruption begins, a current sheet (CS) forms below the rising configuration, which leads to flare reconnection. The flare reconnection is believed to be the major process that releases the magnetic free energy \citep[][]{Forbes1991}.
 
Despite the wide-spread consensus on overall picture given above, there has been a long debate over the roles of multiple physical mechanisms in CME initiation \citep[][]{Torok2023}. The early works on the catastrophe model \citep[e.g.,][]{VanTend1978,Molodenskii1987} focused on the mechanical stability of the pre-eruptive configuration. \citet{Forbes1991} studied the configuration that includes an MFR and a CS. They reproduced the loss of mechanical stability of the configuration in an ideal-MHD regime, although in their model, the absence of flare reconnection prevents a significant release of magnetic energy. The catastrophe model was extended by a number of authors \citep[e.g., ][]{Isenberg1993,Shibata1995,Lin2000,Lin2004}. \cite{Kliem2014} demonstrated that the catastrophe of a torus-shaped MFR is equivalent to the torus instability \citep[e.g., ][]{Bateman1978,Kliem2006,Demoulin2010}. The torus instability theory treats the pre-eruptive MFR as a torus in an external magnetic field $\mathbf{B}_{ex}$, and focuses on the Lorentz self-force (i.e., the hoop force) of the torus and the force due to $\mathbf{B}_{ex}$ (i.e., the strapping force). In the pre-eruptive stage, the two forces balance, and the torus is stable against its expansion in the major radius. When such expansion becomes unstable, the torus instability occurs, which is postulated to initiate the eruption. Another type of instability widely discussed in the CME initiation research is the kink instability \citep[][]{HOOD1981,Torok2005}. The kink instability theory suggests that when the twist of the MFR exceeds a threshold, the MFR becomes unstable against the helical deformation. Both the torus and kink instability theories focus on ideal magnetohydrodynamic (MHD), suggesting that magnetic reconnection is not necessarily required for the initiation of an eruption. Furthermore, in at least the case of the torus instability, the acceleration of the CME could be due to a purely ideal process.

As proposed by \citet{Antiochos1999} \citep[see also][]{Lynch2008}, another physical process, so-called breakout reconnection, can occur when $\mathbf{B}_{ex}$ exhibits a multipolar topology. A multipolar photospheric field distribution will, in general, produce a magnetic null point along with separatrix surfaces in the corona. Consider a stable pre-eruptive configuration below the null-point. If the configuration moves upward due to magnetic stress buildup, the null point will transform into a horizontal CS (i.e., the breakout CS) above the erupting configuration, at which point breakout reconnection can occur. The breakout reconnection cancels the strapping field near the MFR and the background field above the null line, thereby reducing the strapping field and facilitating the eruption. Although breakout reconnection can facilitate the eruption, \citet{DeVore2008} argued that breakout reconnection may also erode the flux of the erupting magnetic configuration. They found that a sufficiently strong background field can confine the eruption after the erosion. \citet{Chen2023a} proposed a model for the confined MFR eruption, in which the breakout reconnection completely erodes the poloidal flux of the MFR. Their model was validated by the observation of a failed solar filament eruption in \citet{Chen2023b}. Another reconnection-driven mechanism for CME initiation is the so-called tether-cutting model \cite{Moore1992,Jiang2021}. In this model, component reconnection (also termed tether-cutting) is assumed to begin inside the filament channel due to the strong gradients in shear there. The eruption onset is postulated to be due to the spontaneous transition of this tether-cutting reconnection from a slow to a fast, flare state \citep{Jiang2021}. 

The key difference between the breakout and the ideal instability mechanisms for eruption is in the topology of the pre-eruptive magnetic field, {\it i.e.}, the filament channel. The ideal models inherently assume a MFR, a twisted flux rope configuration, because they require a net current through the filament channel in order to have either the hoop force for the torus, or the helical bending for the kink. For breakout the net current is not needed, so the initial configuration can be a MFR or a pure CS. Note that once flare reconnection begins, however, a CS will quickly be transformed into a twisted flux rope, so that the actual erupting structure is always a MFR.   

A critical factor that distinguishes the scenarios above and the nature of the eruption, in general, is the background magnetic field. The external magnetic field $\mathbf{B}_{ex}$ of the MFR is composed of the local active region (AR) magnetic field $\mathbf{B}_{AR}$ due to the magnetic poles at the two sides of the PIL, and the background field $\mathbf{B}_{BG}$ due to the sources outside the AR (e.g., a global dipolar field). With a fixed $\mathbf{B}_{AR}$, the variation of $\mathbf{B}_{BG}$ can affect the eruption in the following ways: (1) The orientation of $\mathbf{B}_{BG}$ with respect to $\mathbf{B}_{AR}$ controls whether the coronal null could exist, thus determining whether the breakout reconnection could occur. (2) If the null exists, then the strength of $\mathbf{B}_{BG}$ affects to what extent the breakout reconnection erodes the flux of the erupting MFR. (3) The strength of $\mathbf{B}_{BG}$ also affects the radial modulation of the total external field $\mathbf{B}_{ex}$, thus affecting the threshold for the torus instability. We conclude that it is important to investigate, in detail, how the variation of the background magnetic field affects multiple physical processes of the MFR eruption.

MHD simulation is the primary tool for modeling and studying the physical processes in CMEs. A common strategy for constructing CME simulations is to insert analytical MFR models into the simulation domain. A number of CME simulations \citep[e.g.,][]{Manchester2004,Jin2016,Singh2018,Liu2025} employed the Gibson-Low (GL) MFR model from \citet{Gibson1998} to generate eruptions. In most of these works, the GL MFR is inserted into the pre-existing coronal simulation and erupts shortly after the insertion. Therefore, these studies focus particularly on CME propagation and space weather effects rather than on initiation. Another MFR model, i.e., the Titov-D\'emounlin (TD) model from \citet[][]{TD1999}, is more often used to construct CME simulations that include the pre-eruptive stage. \citet[][]{TD1999} studied the force-equilibrium of the thin force-free toroidal MFR and proposed the analytical TD model for the MFR. \citep[][]{Titov2014} proposed a modified TD (mTD) model for embedding the MFR in the potential magnetic field, which was later used in a number of simulations \citep[e.g.,][]{Torok2018,Downs2021} to initiate MFR eruptions. \citet{Sokolov2023} investigated the non-force-free equilibrium of a $\beta>0$ magnetized plasma and proposed another mTD model for the MFR. When inserting the TD MFR model into the pre-existing coronal simulation, the parameters of the TD MFR are chosen so that the configuration is in a near-equilibrium state, which avoids an immediate eruption and allows for the simulation of a pre-eruptive stage.

Another strategy in CME simulations is to impose photospheric driving at the inner boundary to either build up free magnetic energy from a potential field or to break the equilibrium of an existing pre-eruptive configuration. There are several types of photospheric driving, such as the convergent flow near PIL that generates flux cancellation \citep[e.g.,][]{Linker2003,Torok2018}, the shearing flow near PIL \citep[e.g.,][]{Antiochos1999,Lynch2008,Karpen2012}, flux emergence \citep[e.g.,][]{Fan2016,Fan2017,Torok2024}, and the energization of filament channel through magnetic shear \citep[e.g.,][]{Mikic2018}. \citet{Antiochos2013} argued that small-scale vortical motions in the solar photosphere can statistically inject magnetic helicity into the solar corona, which is ultimately condensed into large-scale magnetic shear through reconnection. Based on this theory, \citet{Dahlin2022} developed the STatistical Injection of Condensed Helicity Model \citep[STITCH,][]{Dahlin2022} as a photospheric driving at the inner boundary of coronal models. The STITCH model introduces a source term into the equations for the horizontal components of the magnetic field, for which the analytical expression is similar to that of the energization model \citep[e.g.,][]{Mikic2018}. 

This work investigates the roles of the background magnetic field and breakout reconnection in initiating MFR eruptions. We use the TD MFR model and the STITCH model to simulate the energy buildup and MFR eruption with different background fields. We examine the eruption properties and the physical processes, including magnetic reconnection and the occurrence of the torus instability, and investigate their dependence on the background magnetic field. Section~\ref{section_methods} describes the MHD simulations. Section~\ref{section_results} presents the simulation results and outstanding features. In Section~\ref{section_mechanism}, we analyze the CME mechanisms of each simulation and discuss the effect of the background field. We discuss the primary conclusions of this work in Section~\ref{section_summary}.

\section{Methods}\label{section_methods}

Our simulations are based on three simple and idealized photospheric radial magnetic field maps ($B_{r}$ maps). Each $B_{r}$ map is generated by two two pairs of magnetic charges beneath the solar photosphere: (1) a local pair of charges centered at $(x,y,z)=(-0.95R_{s},0,0)$, and (2) a background pair of charges centered at the origin. Here $R_{s}$ is the solar radius and the origin of the coordinate system is at the solar center. The two pairs of magnetic charges correspond to $\mathbf{B}_{AR}$ and $\mathbf{B}_{BG}$, respectively. In the following text, the three external magnetic fields correspond to Case 1, Case 2, and Case 3, respectively.

The three cases contain the identical local bipolar field $\mathbf{B}_{AR}$. The local positive and negative magnetic charges are located at $(-0.95R_{s},0,\pm0.1R_{s})$, in other words, the two local magnetic charges are located at a depth of $\approx0.05R_s$ below the surface and oriented along the $z$ axis. We specify the local magnetic charges so that the maximum value of the radial component of $\mathbf{B}_{AR}$ at $r=R_{s}$ is approximately $250$ G. The two magnetic charges corresponding to the background field are located on the z-axis and close to the origin. We use the strength of $\mathbf{B}_{BG}$ at the north pole of the photosphere, $B_{BG}^{N}$, to characterize the strength of the background field. In the three cases, the values of $B_{BG}^{N}$ are $1$ G, $-5$ G and $-10$ G, respectively. The background and local pairs of magnetic charges have the same orientation in Case 1, while they have opposite orientations in Cases 2 and 3.

\begin{figure}[tp!]
\centering {\includegraphics[width=1.0\hsize]{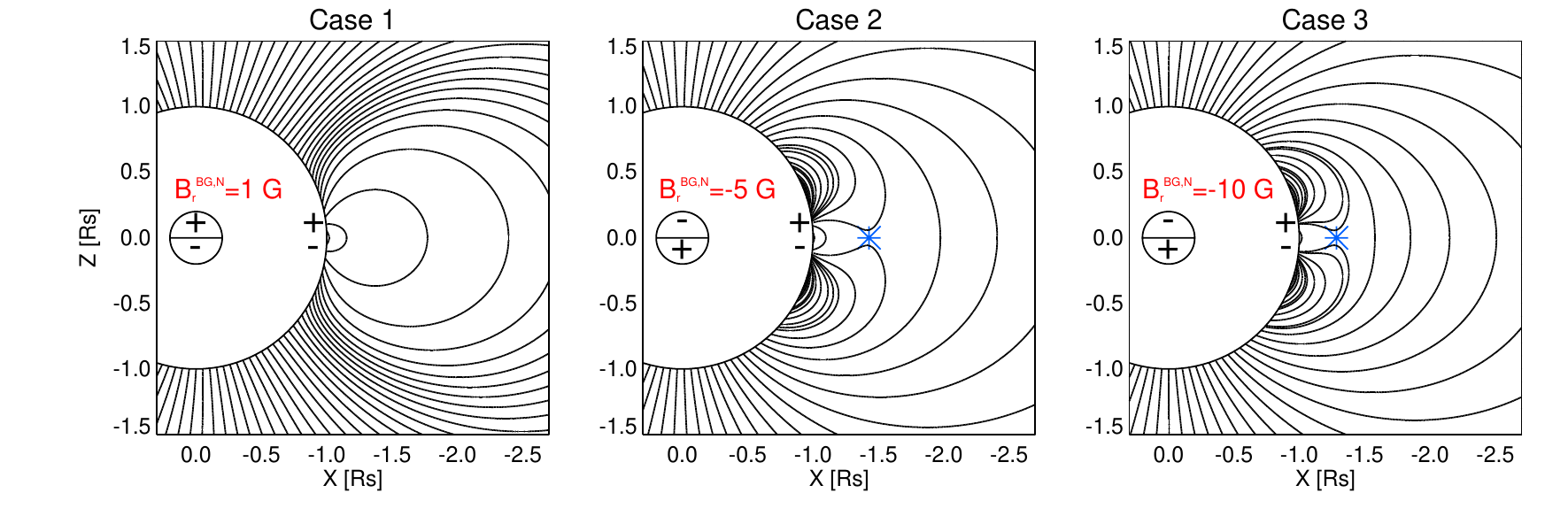}} 
\caption{The magnetic field in the $y=0$ plane in the three cases. The strength of the background dipole is characterized by $B_{r}^{BG,N}$. The null points in Cases 2 and 3 are labeled with blue asterisks.} \label{fig_potential_fields}
\end{figure}

We calculated the magnetic field generated by the two pairs of magnetic charges and traced magnetic field lines to illustrate the external field $B_{ex}$ corresponding to each $B_{r}$ map. Figure~\ref{fig_potential_fields} shows the magnetic field lines in the $y=0$ plane. Note that our field is fully 3D, but $y=0$ is a symmetry plane for the initial magnetic field. $B_{ex}$ in Case 1 exhibits a simple connectivity. Since the local and background pairs of magnetic charges have the same orientation and are both antisymmetric about the $z=0$ plane, the primary photospheric PIL lies at the equator. $B_{ex}$ in Cases 2 and 3 exhibits a multipolar topology due to the opposite orientation of the two pairs of magnetic charges. A coronal null line is present in both cases due to the symmetry of this system in which the local pair of magnetic charges is exactly centered on the equatorial plane and oriented exactly in the vertical direction, as in the studies of \citet{DeVore2008} and \citet{Dahlin2025}. The advantage of this configuration is that we expect the flare current sheet to form along the equatorial plane, allowing accurate determination of the exact onset of flare reconnection \citep{Dahlin2025}. The blue asterisks in Figure~\ref{fig_potential_fields} label the central null point, i.e., the intersection between the null line and the $y=0$ plane. The null point in Case 2 is $\approx1.43R_{s}$ from the solar center, while in Case 3, the null point is only $\approx 1.28R_{s}$ from the solar center due to the stronger background field.

\subsection{Alfv\'en Wave Solar atmosphere Model-Realtime}

We use the Alfv\'en Wave Solar atmosphere Model-Realtime \citep[AWSoM-R, as described by][]{Sokolov2021} to perform the MHD simulations. AWSoM-R is an upgraded version of the Alfv\'en Wave Solar atmosphere Model \citep[AWSoM,][]{vanderholst:2014}. Similar to AWSoM, AWSoM-R calculates the effects of several physical processes, e.g., the electron heat conduction, radiative cooling heat loss, and the Alfv\'en wave coronal heating \citep[see][for more details]{vanderholst:2014}.

Both AWSoM and AWSoM-R employ a spherical computational domain with the inner boundary in the upper chromosphere at $r\approx1\,R_s$ and the outer boundary set in the outer solar corona at $r\approx24\, R_s$. The important distinction between AWSoM-R and AWSoM is in introducing the threaded-field-line model (TFLM). The simulation domain of AWSoM-R is divided by a spherical surface at $r=R_{b}$ ($1.01\,R_{s}<R_{b}<1.1\,R_{s}$) into two spherical shells. In the outer shell, AWSoM-R solves the same three-dimensional (3D) extended MHD equations as AWSoM. The inner spherical shell extends from $R_{s}$ to $R_{b}$, which includes the transition region and the upper chromosphere. In this region, AWSoM-R uses TFLM, which solves a set of 1-D problems for plasma motion along magnetic field lines (\textit{threads}) rather than solving the 3-D MHD equations. Hereafter, the two regions are referred to as the 3D MHD region and the TFLM region, respectively.

The TFLM region contains a set of 1D magnetic threads that connect the 3-D MHD region to the inner boundary at $r=R_{s}$. The magnetic field in the TFLM region is assumed to be the potential field source surface (PFSS) solution with the given magnetic map applied as the inner boundary condition. The threaded field lines are traced along the magnetic field of the PFSS solution, starting from the centers of 3D grid cells near the $r=R_{b}$ surface to the inner boundary at $r=R_{s}$. The step size ($ds$) we use for the field line tracing is $10^{-4}R_{s}\approx70\,\mathrm{km}$. The velocity in the TFLM region is field-aligned, i.e., the velocity vector is parallel to the magnetic field vector. In this way, we reduce the 3D MHD equations to the field-aligned hydrodynamic (HD) equations. A detailed description of TFLM is given in \citet{Sokolov2021}. The TFLM solution provides the boundary conditions of temperature, density, velocity, and the radial magnetic field for the 3D MHD region.

\begin{figure}[tp!]
\centering {\includegraphics[width=1.0\hsize]{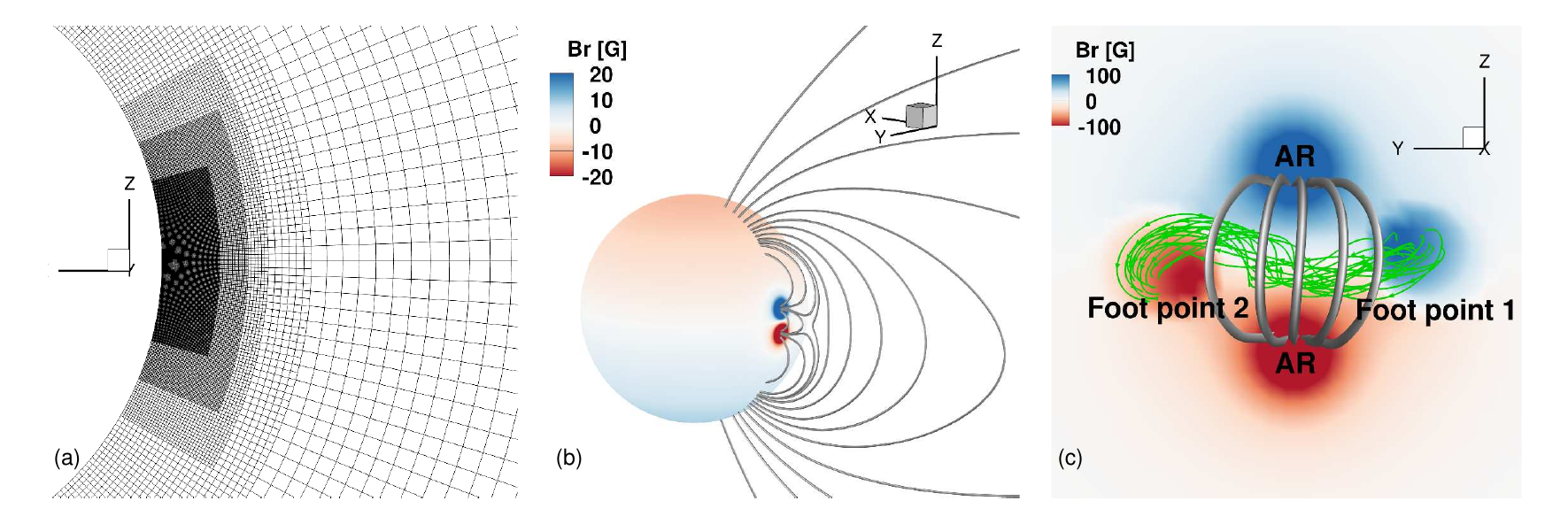}} 
\caption{Simulation setup. (a): The simulation grid in the $y=0$ plane. (b): The simulated magnetic field in Case 3 before the insertion of the MFR. (c): The AR and the MFR (green field lines) in Case 3.}\label{fig_setup}
\end{figure}

The aim of using TFLMs is to achieve a higher numerical performance in real-time simulations. In simulations with hyperbolic equations and global time-stepping, the time step $dt$ is restricted by Courant–Friedrichs–Lewy (CFL) condition \citep[][]{CFL1928}, i.e., $dt\leq\mathrm{min}\left(\mathrm{CFL}\cdot\frac{dx}{V_{\mathrm{max}}}\right)$. Here $\mathrm{CFL}$ is of order unity, $dx$ represents the local grid size, and $V_{\mathrm{max}}$ is the maximum characteristic wave speed in the direction of $dx$. 
Since both AWSoM and AWSoM-R use an extremely small grid size to resolve the steep gradients in temperature and density in the transition region, the time step is controlled by the CFL condition near the inner boundary in both models. However, AWSoM and AWSoM-R have different expressions for $V_{\mathrm{max}}$. AWSoM solves the 3D MHD equations in the entire domain, including the region from $R_{s}$ to $R_{b}$. Therefore, the CFL condition in AWSoM yields $dt\leq\mathrm{min}\left(\mathrm{CFL}\cdot\frac{dr}{V_{\mathrm{fast}}}\right)$, where $V_{\mathrm{fast}}$ is the fast-mode magnetosonic wave speed of the MHD equation. 
In AWSoM-R,  the CFL condition of the TFLM yields $dt\leq\mathrm{min}\left(\mathrm{CFL}\cdot\frac{ds}{V_{\mathrm{sound}}}\right)$, where $V_{\mathrm{sound}}$ is the sound speed. If one applies $ds=dr$ and an identical $\mathrm{CFL}$ for AWSoM and AWSoM-R, then the time steps of the two models would yield a ratio of $\frac{V_{fast}}{V_{sound}}$. This ratio is significantly larger than unity in the regions with low plasma-$\beta$. Therefore, AWSoM-R achieves higher performance in real-time simulations than AWSoM. However, we note that the use of TFLMs trades off the accuracy of dynamics within $R_{s}<r<R_{b}$. AWSoM-R simulations only capture the magnetic evolution above the $r=R_{b}$ surface, as the magnetic field below this surface is assumed to be the PFSS solution. In this work, we use $R_{b}=1.01R_{s}$, which ensures that the impact of TFLMs on the dynamics of the eruption is negligible.

\begin{table*}[ht!]
\caption{Iteration, region, and the resulting angular resolution of each refinement.} \label{tab_amr}
\setlength{\tabcolsep}{0.25cm}
\begin{tabular}{c|cc}
\hline\hline
Iteration & Refinement Region $(r,\phi,\lambda)$ & Resulting Angular resolution \\ \hline
12,000 & $[1.01\; R_s,1.4\; R_s]\times[150^{\circ},210^{\circ}]\times[-30^{\circ},30^{\circ}]$ & $0.938^{\circ}$ \\
24,000 & $[1.01\; R_s,1.3\; R_s]\times[155^{\circ},205^{\circ}]\times[-25^{\circ},25^{\circ}]$ & $0.469^{\circ}$ \\
36,000 & $[1.01\; R_s,1.25\; R_s]\times[160^{\circ},200^{\circ}]\times[-20^{\circ},20^{\circ}]$ & $0.234^{\circ}$ \\
48,000 & $[1.01\; R_s,1.2\; R_s]\times[165^{\circ},195^{\circ}]\times[-15^{\circ},15^{\circ}]$ & $0.117^{\circ}$ \\
\hline
\end{tabular}
\end{table*}

We first used AWSoM-R to construct the steady-state external magnetic field of each case. Before each simulation, we used the iterative Finite-DIfference Potential Field Solver \citep[FDIPS,][]{toth2011} to derive the PFSS solution of the $B_{r}$ map. We then used the PFSS solution to derive the grid of each TFLM and set the initial magnetic field in the 3D MHD region. We use the adaptive mesh refinement \citep[AMR,][]{gombosi2003adaptive, toth2012adaptive} to decompose the 3D domain above $R_{b}=1.01R_{s}$ into blocks with $6\times8\times8$
cells in the $(r,\phi,\lambda)$ directions. Here $\phi$ and $\lambda$ denote the longitude and latitude, respectively. At the onset of each simulation, the angular grid size is $1.875^{\circ}$ above $r=1.1\; R_s$ and $0.938^{\circ}$ below $r=1.1\; R_s$. We perform the refinement after every 12,000 iterations. Table~\ref{tab_amr} shows the region and the resulting resolution of each refinement. Figure~\ref{fig_setup} (a) presents the grid in the $y=0$ plane after the four refinements. The resulting angular grid size at the center of the AR is $\approx0.118^{\circ}$. We performed 60,000 iterations with the local time stepping to derive the steady-state coronal magnetic field without the MFR. Figure~\ref{fig_setup} (b) shows the result for Case 3 as an example. We note that the steady-state external field of each case has the same topology as the field shown in Figure~\ref{fig_potential_fields}, except for the addition of an open field region near the two poles. 

\subsection{Titov-D\'emoulin Flux Rope Model}

We apply the TD-type eruptive event generator developed by \citet[][]{Sokolov2023} to build the pre-eruptive magnetic configuration. The original TD flux rope model was proposed by \citet{TD1999}. They studied the force-equilibrium of a thin force-free toroidal magnetic flux rope (MFR) such that the minor radius $a$ is much smaller than the major radius $R_{0}$, i.e., $a\ll R_0$. In the absence of an external field, such a configuration would be disrupted by the hoop force. To prevent this, a \textit{strapping magnetic field} was included in the model created by a pair of magnetic point charges located at the axis of the MFR. An analytical formula was derived to build the 3D magnetic field of such a MFR. \citet{Sokolov2023} investigated the non-force-free equilibrium of a $\beta>0$ magnetized plasma and proposed an mTD model for the MFR.

\begin{table*}[ht!]
\caption{mTD parameters for each model.} \label{tab_td}
\setlength{\tabcolsep}{0.4cm}
\begin{tabular}{c|ccccc}
\hline\hline
Case & $R_{0}$ & $a$ & $B_{ex}$ [G] & $|I^{tot}|$ [A] & $E_{B,\mathrm{TD}}$ [erg]\\ \hline
Case 1 & \multirow{3}{*}{$0.16R_{s}$} & \multirow{3}{*}{$0.06R_{s}$} & $14.40$ & $8.895\times10^{11}$ & $4.127\times10^{32}$\\
Case 2 & & & $14.13$ & $8.730\times10^{11}$ & $3.976\times10^{32}$\\
Case 3 & & & $13.02$ & $8.044\times10^{11}$ & $3.376\times10^{32}$\\
\hline
\end{tabular}
\end{table*}

Once the steady-state solution for the solar corona was obtained, we inserted the mTD MFR into the AR. Figure~\ref{fig_setup} (c) shows the AR and the MFR in Case 3 as an example. We label the magnetic poles of the AR and the MFR footpoints. The axis of the inserted MFR passes through the two monopoles of the AR located at $\left[-0.95R_{s},0,\pm0.1R_{s}\right]$. The major and minor radii of the configuration are $0.16R_{s}$ and $0.06R_{s}$, respectively. The total toroidal current ($I^{tot}$) and the strength of the toroidal magnetic field of the MFR are chosen to guarantee that the resulting configuration is near mechanical equilibrium. According to \cite{Sokolov2023}, the condition for an MFR to be radially balanced in a uniform external field $\mathbf{B}_{ex}$ is that the hoop force of the MFR equals the Lorentz force due to $\mathbf{B}_{ex}$:
\begin{equation}
    \left[B_{ex,a}+\frac{L_{0}I^{tot}}{4\pi (R_{0}^{2}-a^{2})}\right]I^{tot}=0\label{balance_condition}
\end{equation}
Here $L_{0}$ is the self-induction coefficient of the MFR determined by the geometry. $B_{ex,a}$ indicates the axial component of $\mathbf{B}_{ex}$. The condition above ensures that $I^{tot}$ is proportional to $\mathbf{B}_{ex}$ for a given geometry. Since $\mathbf{B}_{ex}$ is non-uniform in our models, we estimated $B_{ex,a}$ by the magnetic field at the apex of the mTD flux rope. Table~\ref{tab_td} shows the parameters of the mTD flux rope of each case. The value of $B_{ex}$ varies for $\approx 10\%$ among the three models. We also calculated the magnetic energy of the inserted mTD MFR ($E_{B,\mathrm{TD}}$) for each case, which yields a relative difference of $\approx 22\%$. Note that $E_{B,\mathrm{TD}}$ includes both the magnetic free energy and the potential field energy associated with the MFR footpoints.

\begin{figure}[tp!]
\centering {\includegraphics[width=1.0\hsize]{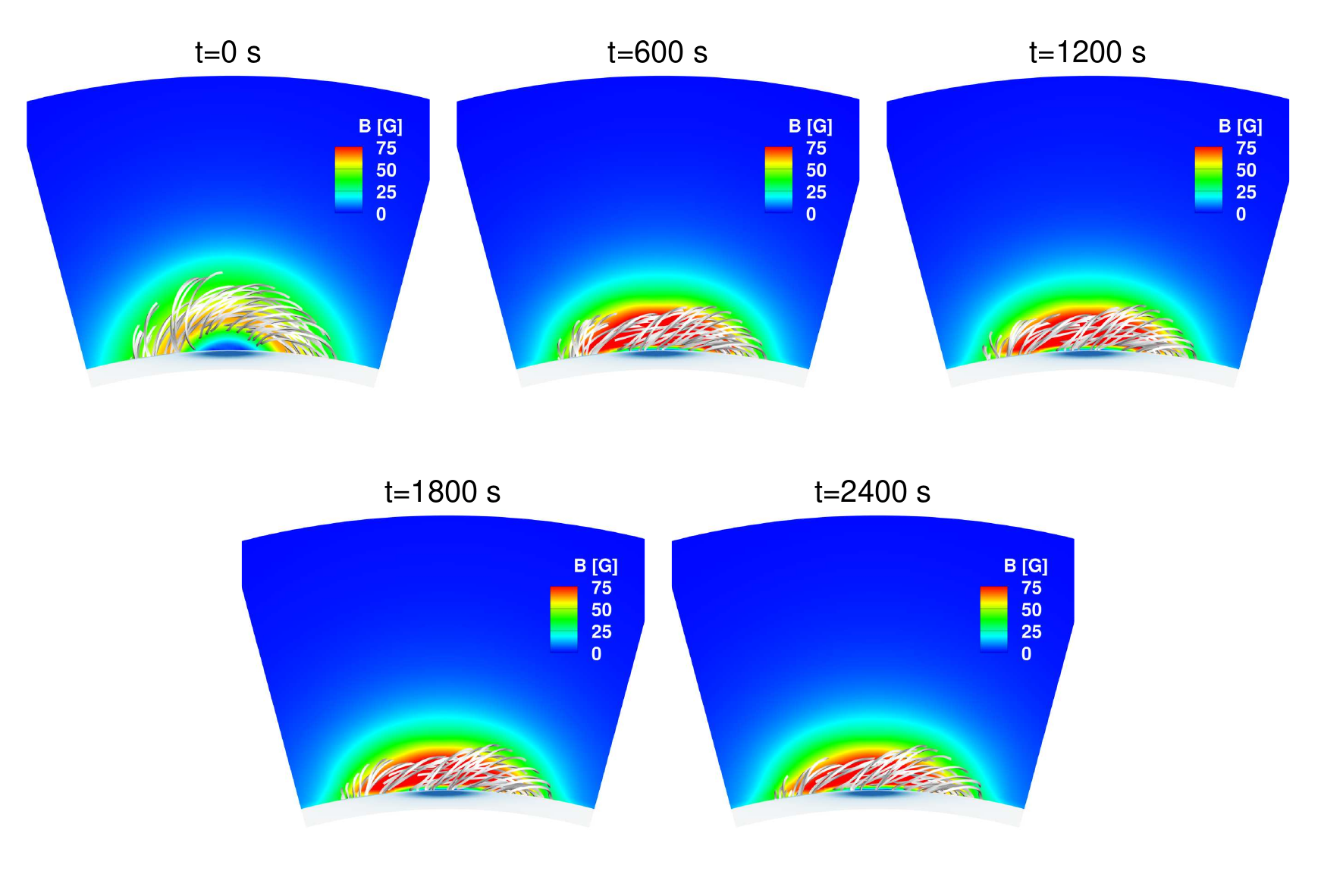}} 
\caption{The MFR viewed from $+z$-axis during the relaxation from $t=0$s to $t=2400$ s in Case 1.}\label{fig_relaxation}
\end{figure}

Upon inserting the MFR, we switch off the local time stepping scheme and continue the MHD simulations with the global time stepping (i.e., time-accurate) scheme. The moment of MFR insertion is denoted as $t=0$. We note that the insertion of mTD MFR also modifies the magnetic field in the TFLM region. Therefore, we adopted a new set of TFLMs after the insertion. In the first $40$ minutes of the simulation, we perform a MHD relaxation of the MFR to obtain a steady-state configuration. Figure~\ref{fig_relaxation} shows the MFR at five moments during the relaxation in Case 1. The appearance of the MFR shown in Figure~\ref{fig_relaxation} is similar in the three cases. The MFR moves downward from $t=0$ s to $t=600$ s, while from $t=600$ s to $t=2400$ s, neither the magnetic field lines nor the distribution of the magnetic field strength exhibits a significant change. This indicates a relaxation process of $40$ min is enough for the MFR to approach a quasi-equilibrium state. Note also that there is no evidence of a writhing motion, indicating that the MFR is not susceptible to the kink instability. This is expected from Figure~\ref{fig_setup} (c), which shows that the twist of the MFR is only one turn or less.

\subsection{STatistical Injection of Condensed Helicity Model}

To trigger the eruption, we apply the STITCH Model from \citet{Dahlin2022} to the MFR. The implementation of the STITCH model in the SWMF is described by \citet{vanderHolst2025}. The STITCH model involves a source term, $\frac{\delta \mathbf{B}_{s}}{\delta t}$, to the horizontal components of the induction equation for magnetic field, near the inner boundary:
\begin{equation}
    \frac{\delta \mathbf{B}_{s}}{\delta t}=\nabla_{s}\times\nabla_{r}\left(\zeta B_{r}\right). \label{stitch_formula}
\end{equation}
Here, $\nabla_{s}$ and $\nabla_{r}$ indicate the horizontal and radial derivatives, respectively. $\zeta$ characterizes the subgrid-scale vortical motions and controls the helicity injection rate. The derivation of Equation~\ref{stitch_formula} can be found in \citet{Mackay2014}. We apply Equation~\ref{stitch_formula} in the region $\left(\phi,\lambda\right)\in\left[180^{\circ}-\Delta\phi,180^{\circ}+\Delta\phi\right]\times\left[-\Delta\lambda,\Delta\lambda\right]$, where $\Delta\phi=10^{\circ}$ and $\Delta\lambda=15^{\circ}$. As in \citet{vanderHolst2025}, the modulation of $\zeta$ within this region is given by
\begin{equation}
    \zeta\left(\phi,\lambda\right)=\zeta_{0}\cos{\left(\frac{\pi}{2}\frac{\phi-180^{\circ}}{\Delta\phi}\right)}\cos{\left(\frac{\pi}{2}\frac{\lambda}{\Delta\lambda}\right)}
\end{equation}
By applying Equation~\ref{stitch_formula} in the simulation, the initially stable MFR is driven to deviate from the equilibrium state, which results in eruption when the stability of the configuration breaks. Note that the STITCH term preserves the radial flux distribution at the photosphere and injects only a pure shear magnetic-field component in the corona at the PIL. Consequently, it acts to increase the axial component of the MFR; thereby, stabilizing it further against a possible kink instability. The eventual destabilization of the system, therefore, must correspond to either a torus-like ideal instability or the effect of breakout or tether-cutting-like reconnection. The duration of the driving process required to trigger the eruption depends on $\zeta_{0}$. In \citet{Dahlin2022}, $\zeta_{0}$ was assumed to be $1.4\times10^{6}$ km$^{2}$s$^{-1}$. We found that $\zeta_{0}=1.4\times10^{6}$ km$^{2}$s$^{-1}$ results in a short driving time of $\simeq 10$ min in our simulations. To achieve a relatively smooth driving and ensure that the flux rope is in a quasi-static state during the driving, we used a smaller value for $\zeta_{0}$, $0.4\times10^{6}$ km$^{2}$s$^{-1}$ in each simulation. The STITCH source term given by Equation~\ref{stitch_formula} is switched on at $t=2400$ s.
We switch off the source term at $t=7200$ s, $5800$ s, and $5200$ s for the three cases, respectively; the reason for this will be given later.

\section{Results} \label{section_results}

\subsection{Magnetic Field Evolution}

\begin{figure}[tp!]
\centering {\includegraphics[width=1.0\hsize]{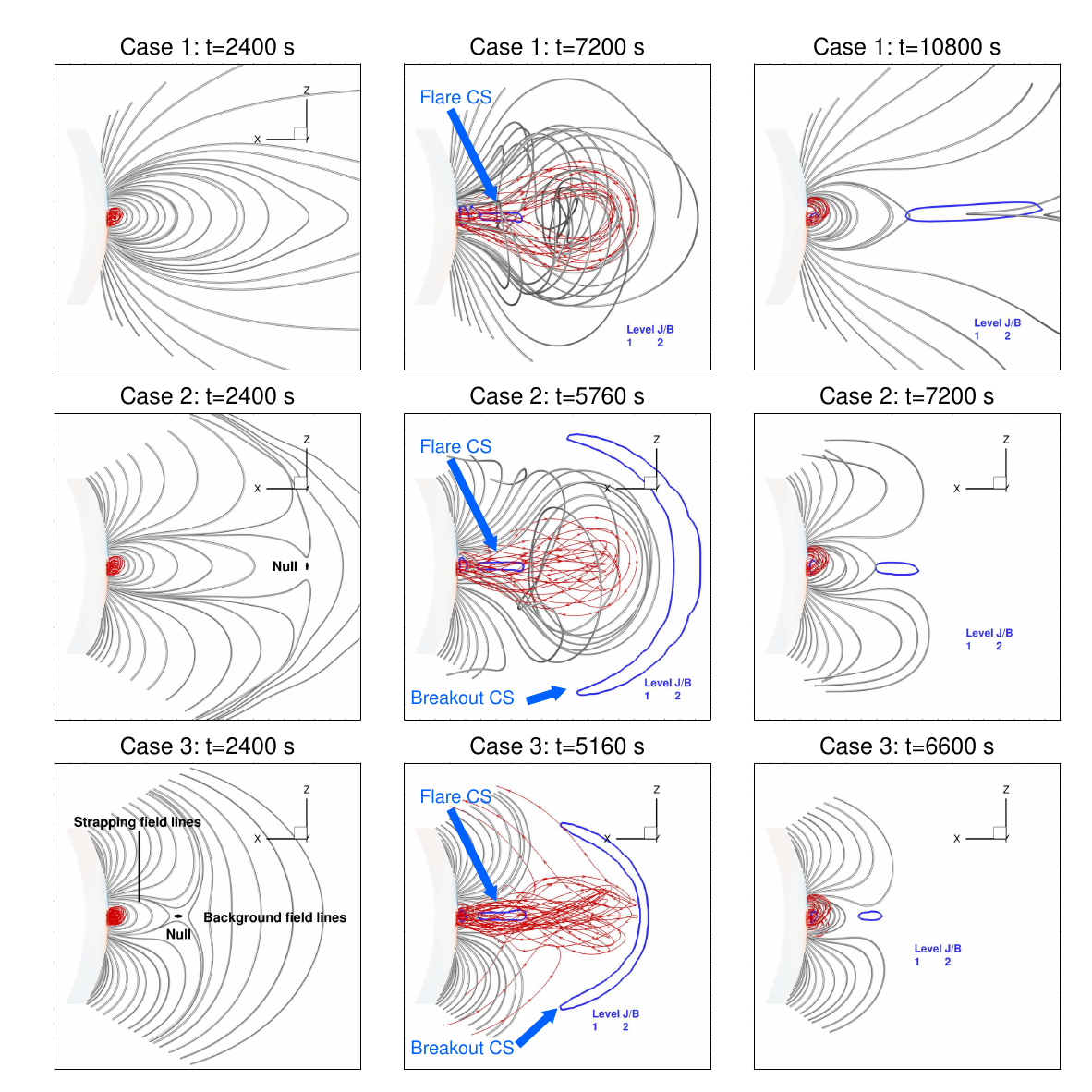}} 
\caption{The simulation results for the three cases. Each row shows the magnetic field of one case at three different moments: (1) at the onset of driving, (2) near the end of driving, and (3) after the eruption. The red field lines are anchored to the footpoints of the MFR. The white rods indicate the field lines surrounding the configuration. The blue contours indicate $\frac{J}{B}=2$ \textmu Am$^{-2}$G$^{-1}\approx\frac{17.5}{\mu_{0}R_{s}}$, which show the positions of current sheets.} \label{fig_results}
\end{figure}

For each simulation, we saved the 3D distribution of the variables with a cadence of $120$ s. Figure~\ref{fig_results} shows the magnetic field in each simulation at three different times corresponding to roughly eruption onset, eruption main phase, and the late relaxation phase. We use contours of the current density over magnetic field strength ($J/B$) in the $y=0$ plane to visualize the current sheets. In Figure~\ref{fig_results} (a), the magnetic field above the MFR shows a helmet-streamer-like morphology, but with a closed field region that is stretched outward due to the presence of the low-lying MFR. The MFR rises and expands after the eruption is triggered. Figure~\ref{fig_results} (b) shows the flare CS that has formed at $t=7200$ s for Case 1. The flare reconnection produces post-reconnection loops with short lengths near the photosphere and adds twisted flux to the MFR. At $t=10800$ s, the magnetic field lines involved in the flare reconnection have become low-lying closed flare loops with low magnetic free energy. The flare CS has been radially elongated at this time. We expect this flare CS to eventually reform the CS of a new helmet streamer if the simulation continues.

\begin{figure}[tp!]
\centering {\includegraphics[width=0.8\hsize]{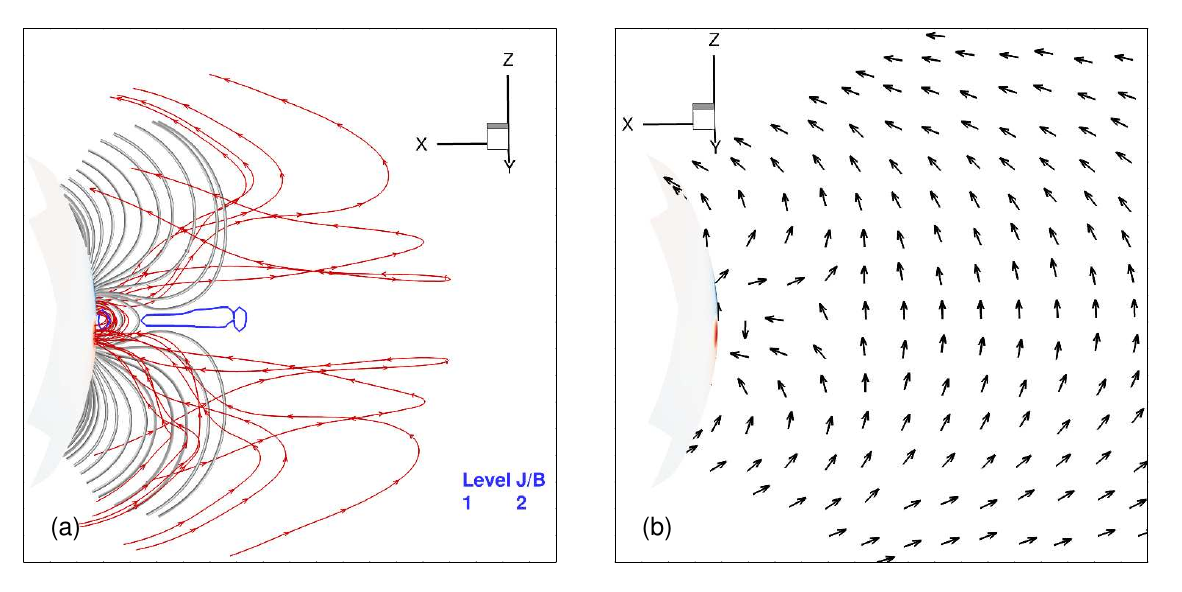}} 
\caption{The simulation results of Case 3 at $t= 5400$ s. (a): The field lines anchored to the footpoints of the initial MFR (red) and surrounding field lines (white). (b): The projected magnetic field vectors in the $y=0$ plane.} \label{fig_failed}
\end{figure}

\begin{figure}[tp!]
\centering {\includegraphics[width=0.9\hsize]{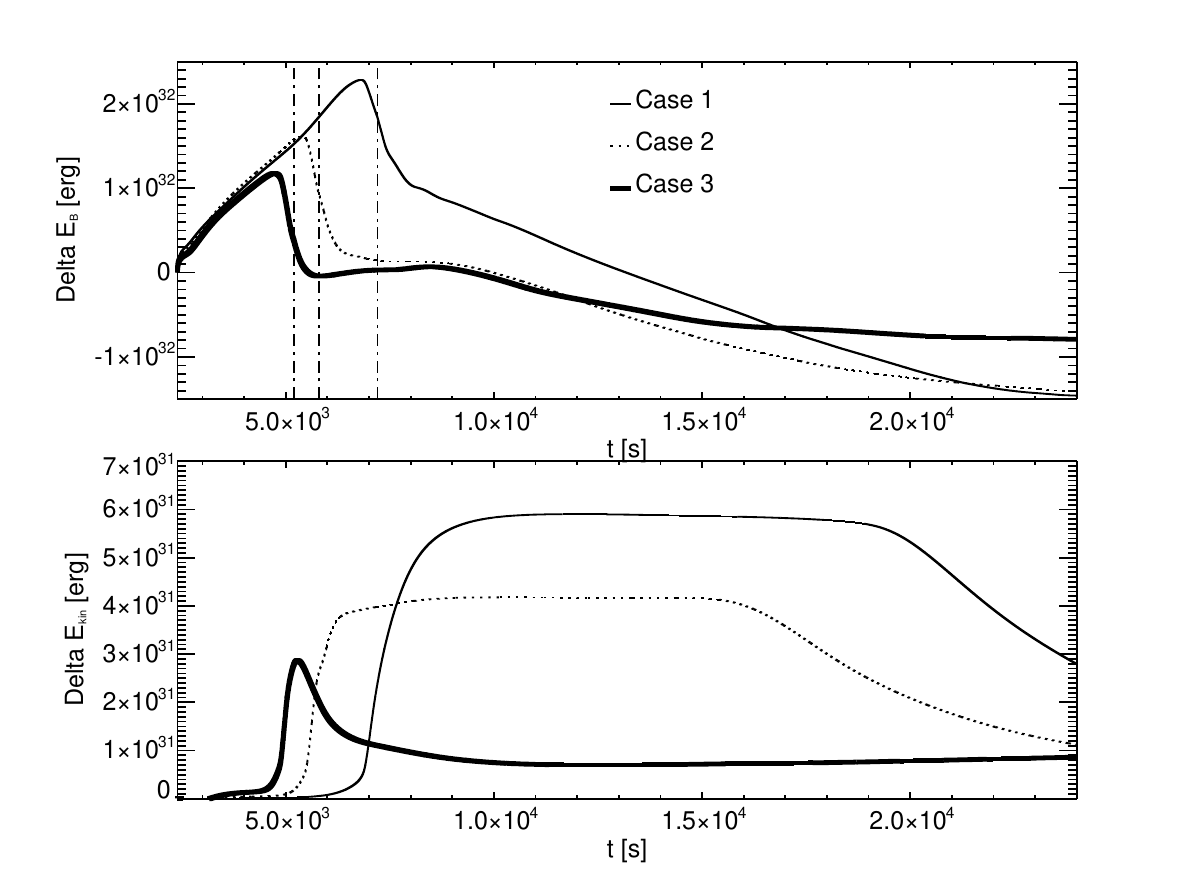}} 
\caption{The evolution of $\Delta E_{B}$ (upper panel) and $\Delta E_{kin}$ (lower panel) in each case. The vertical dashed-dotted lines indicate the end time of STITCH in each simulation.
} \label{fig_energy_curves}
\end{figure}

The initial magnetic field in either of Cases 2 and 3 exhibits a multipolar topology, similar to the corresponding potential field shown in Figure~\ref{fig_potential_fields}. The field lines below the magnetic null connect to the AR, while the field lines above the magnetic null connect to the polar regions. In Figure~\ref{fig_results} (e), we see two CSs at $t=5760$ s: (1) the flare CS, which is similar to that of the first case, and (2) the breakout CS, which evolves from the initial magnetic null. The breakout reconnection cancels AR magnetic flux and overlying magnetic flux. The flare CS of Case 2 at $t=7200$ s is relatively short compared to the radially elongated flare CS in Case 1. We found that the flare CS in Case 2 has already become a new null point at the end of the simulation ($t=10800$ s, which is not presented in Figure~\ref{fig_results}).

Figure~\ref{fig_results} also shows that the pre-eruptive and post-eruptive magnetic fields of Case 3 are similar to those of Case 2. The magnetic field of Case 3 during the eruption also contains both the flare CS and breakout CS. However, we note that at $t=5160$ s (Figure~\ref{fig_results} (h)), the magnetic field of the MFR has been partially canceled with the overlying magnetic field by the breakout reconnection and connected to the high latitude regions. This feature strongly suggests that the breakout reconnection may completely confine the MFR during the simulation. Our examination of the output data shows that the magnetic flux rope has been completely eroded due to the reconnection by the next saved time frame at $t=5280$ s. 
To illustrate this result clearly, we present the magnetic field at $t=5400$ s in Figure~\ref{fig_failed}. Figure~\ref{fig_failed} (a) shows that the direct magnetic connectivity between the two MFR footpoints is broken. The $J/B$ contour corresponding to the breakout CS is not visible, indicating that the breakout reconnection is weak or has terminated. Figure~\ref{fig_failed} (b) shows the projected magnetic field vector in the $y=0$ plane, in which the twisted component of the MFR is absent. Based on the analysis above, we conclude that Case 3 produces a confined flare, or failed eruption.

\subsection{Energetics}

During the simulations, we calculated the total magnetic energy ($E_{B}=\int\frac{B^{2}}{2\mu_{0}}dV$) and the total kinetic energy ($E_{kin}=\int\frac{\rho |\mathbf{v}|^{2}}{2}dV$) at each iteration. Figure~\ref{fig_energy_curves} shows the change in $E_{B}(t)$ and $E_{kin}(t)$ starting from the onset of the STITCH driving ($t=2400$ s). The total magnetic energy first experiences a steady near-linear increase in each case due to the constant injection of shear flux by the STITCH term. Since the STITCH driving is essentially identical in the three cases, the initial rise is very similar, but at some point we note the onset of a sharp drop in the magnetic energy. This is due to the onset of the eruption during which there is a rapid conversion of magnetic energy to kinetic energy. The times of peak $E_{B}$ are $t=6805$ s, $5377$ s, and $4722$ s for the three cases, respectively. We switch off the STITCH driving at $t=7200$ s, $5800$ s, and $5200$ s respectively in the three cases, as indicated by the dashed-dotted vertical lines in Figure~\ref{fig_energy_curves}. We selected these times to occur shortly after the eruption onset in order to minimize further impact by the driving on the eruption energetics.

In the lower panel of Figure~\ref{fig_energy_curves}, the $E_{kin}$ of both Cases 1 and 2 increases first monotonically with time and levels off at some high value. During the late stage of these two Cases, $E_{kin}$ decreases since the MFR propagates through the outer boundary of the domain at $t=24 R_{S}$. However, the $E_{kin}(t)$ profile of Case 3 exhibits a markedly different trend in that $E_{kin}(t)$ peaks at $t=5278$ s and then decays rapidly. The decrease of $E_{kin}(t)$ after the peak time indicates the confinement of the eruption. The peak time of $E_{kin}(t)$ matches the time when the MFR disappears (between $t=5160$ s and $t=5280$ s). The high density carried by the MFR is still present immediately after the MFR disappears by reconnection with the background field. As the CME mass propagates and slows down, the kinetic energy gradually decreases. Note that the $E_{B}(t)$ profile of Case 3 experiences a weak increasing phase after $t\approx 5700$ s and reaches a second peak at $t\approx 8500$ s, due to the kinetic energy being partially converted back to magnetic energy.

Several important points should be made regarding Fig.~\ref{fig_energy_curves}. First, note that the initial magnetic energy value of zero in the figure does not correspond to zero free energy, because the initial state includes the mTD flux rope. It is only the state before the turn on of STITCH driving. This explains why the magnetic energies of all three cases drop below zero in the figure. A key point is that Cases 2 and 3 appear to be more explosive than Case 1 in that the drop in magnetic energy and the rise in kinetic energy right after the magnetic energy peak are sharper. Furthermore, during the later stage of the simulation, the magnetic energy in Case 1 decays significantly faster than Cases 2 and 3, which is likely due to an extended phase of reconnection required to rebuild the high-lying helmet streamer system. We believe that the explosiveness of Cases 2 and 3 is due to the feedback between the breakout and flare reconnection as discussed for example by \citet{Karpen2012}. Note also that the stronger background field of Case 3 leads to an earlier onset than in Case 2, and to similar dynamics, except that no CME escapes. These results have important space weather implications. They suggest that the strength of an eruptive event, the total energy released, can be forecast by monitoring the properties of the filament channel, but predicting whether a CME occurs or not requires measuring the background field.   

\subsection{MFR Dynamics}

\begin{figure}[tp!]
\centering {\includegraphics[width=1.0\hsize]{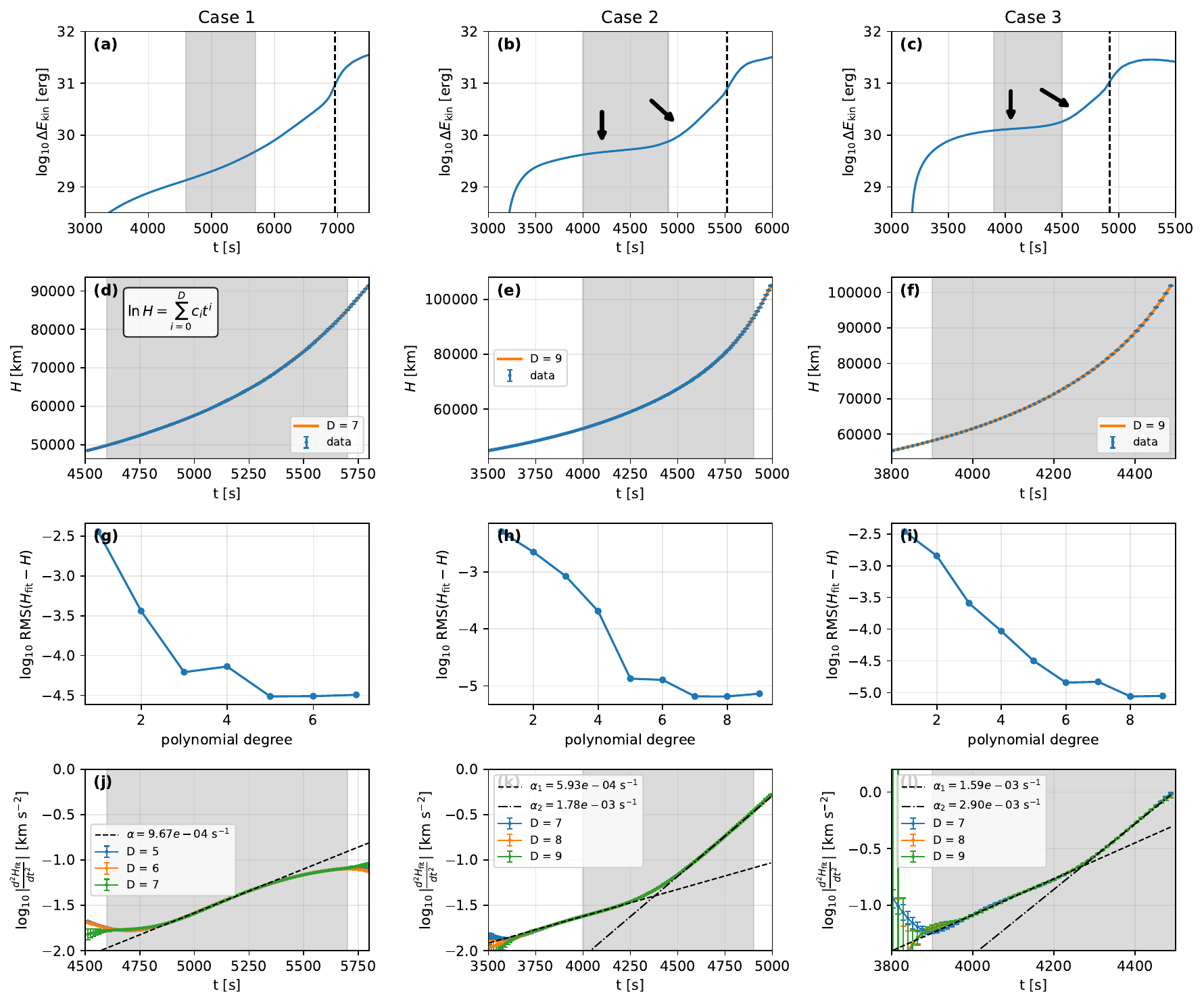}} 
\caption{(a)-(c): The kinetic energy on a logarithmic scale. The vertical dashed line labels the onset of flare reconnection. (d)-(f): The blue points represent the MFR apex height ($H$) as a function of time. The orange curve represent $H_{\mathrm{fit}}(t)$ derived using polynomial fit.(g)-(i): The RMS of $\left|H_{\mathrm{fit}}-H\right|$ using different values of $D$. (j)-(l): The acceleration on a logarithmic scale analytically derived from $H_{\mathrm{fit}}(t)$ using the highest three polynomial degrees. The black dashed and dotted-dashed lines indicate linear fits of $\log{\left|\frac{d^{2}H_{\mathrm{fit}}}{dt^{2}}\right|}(t)$ over several intervals.} \label{fig_speed}
\end{figure}

To study the acceleration of the MFR, we present $\Delta E_{kin}$ on a logarithmic scale in the first row of Figure~\ref{fig_speed}. We use the dotted-dashed line in each panel to indicate the onset of flare reconnection. The determination of the onset will be given in the next section. The evolution of the kinetic energy experiences a transition near the flare reconnection onset, as $\log{\Delta E_{kin}(t)}$ transitions from a steady growth to a more rapid increase. An interesting feature is that in Cases 2 and 3, the evolution of the kinetic energy before the flare reconnection transitions from a stage in which $\Delta E_{kin}(t)$ nearly saturates to a steady growth stage, as is pointed out by the black arrows and the gray regions in panels (b) and (c). However, this transition is not present in Case 1. The gray region in panel (a) only shows a weak change in the energy growth rate. This comparison suggests that Cases 2 and 3 likely include a different MFR acceleration process than Case 1. 

We chose the $x$-axis corresponding to the radial direction to analyze the dynamics of the MFR apex. We first saved the distribution of MHD variables along the $x$-axis with a cadence of $12$ s. Then we determined the MFR apex height $H$ at each moment by searching for the position that is both inside the MFR and the $z$ component (i.e., in the $\theta$ direction) of the magnetic field vanishes there. In the second row of Figure~\ref{fig_speed}, the blue dots represent the $H(t)$ data we derived from the simulation  with error bars. The error here is introduced by the discretization of $B_{z}$ when determining $H$. In order to reduce the effect of this error on our following analysis of MFR acceleration, we fit $\ln{H(t)}$ of each case using a polynomial function:
\begin{equation}
    \ln{H_{\mathrm{fit}}(t)}=\sum_{i=0}^{D}c_{i}t^{i}
\end{equation}
where $D$ is the degree of polynomial and $c_{i}$ represents the coefficients. We apply the polynomial fit to a time interval slightly wider than the gray region of each case. The orange curves in Figure~\ref{fig_speed} (d)-(f) show the results of polynomial fits using $D=\left[7,9,9\right]$ for three cases, respectively. The third row of Figure~\ref{fig_speed} shows the root mean square (RMS) of the fit residual ($|H_{\mathrm{fit}}-H|$) using different values of $D$. With $D=\left[7,9,9\right]$ for three cases respectively, the RMS residual is below $10^{-4.5} R_{s}$. We did not use $D\geq10$ to prevent overfitting. We then analytically derived the acceleration $\frac{d^{2}H_{\mathrm{fit}}}{dt^{2}}$ from the polynomial fit, which is shown in the last row of Figure~\ref{fig_speed} on a logarithmic scale. The error bar represents the total error of $\log{\left|\frac{d^{2}H_{\mathrm{fit}}}{dt^{2}}\right|}$ due to both the discretization when determining $H$ and the polynomial fit. For each case, we provide the result using the three highest polynomial degrees. Due to the convergence of the fitting result and the scale of error bars, we conclude that the error of the derived $\log{\left|\frac{d^{2}H_{\mathrm{fit}}}{dt^{2}}\right|}$ can be ignored except for the edges of the fitting interval.

The acceleration curves exhibit a significant difference between Case 1 and the other two cases. In Case 1, we notice a linear growth $\log{\left|\frac{d^{2}H_{\mathrm{fit}}}{dt^{2}}\right|}$ with time, which indicates an exponential growth of acceleration, followed by a decrease in the growth rate after $t=5300$ s. In Cases 2 and 3, however, the acceleration exhibits a clear transition from a relatively slow exponential growth to a rapid exponential growth. To quantify these features, we applied the linear fit to $\log{\left|\frac{d^{2}H_{\mathrm{fit}}}{dt^{2}}\right|} (t)$ with $\log{\left|\frac{d^{2}H_{\mathrm{fit}}}{dt^{2}}\right|}=\alpha t+b$, where $\alpha$ indicates the exponential growth rate of acceleration. For Case 1, we only fit $\log{\left|\frac{d^{2}H_{\mathrm{fit}}}{dt^{2}}\right|}$ over the interval $t\in[4900,5300]$ s, which gives $\alpha=9.67\times10^{-4}$ s$^{-1}$. For either Case 2 or 3, we applied the linear fit to two intervals and derived the growth rates before ($\alpha_{1}$) and after ($\alpha_{2}$) the transition, as shown in the last row of Figure~\ref{fig_speed}. The results suggest that during the transition phase, the relative increase in growth rate is $200\%$ and $82\%$ for Cases 2 and 3, respectively, which is remarkably different from the decreasing growth rate in Case 1.

We point out that these features in the acceleration curves match the evolution of kinetic energy for the three cases. For Case 1, the kinetic energy exhibits a relatively steady growth before the flare reconnection, and $\log{\left|\frac{d^{2}H_{\mathrm{fit}}}{dt^{2}}\right|}(t)$ curve does not exhibit a clear transition. In both Cases 2 and 3, the acceleration experiences a transition from slow growth phase to a fast growth phase. Comparing the first and last rows of Figure~\ref{fig_speed}, we notice that the transition in $\log{\left|\frac{d^{2}H_{\mathrm{fit}}}{dt^{2}}\right|}(t)$ occurs slightly earlier than the transition in $E_{kin}$, which suggests that the relatively rapid growth of acceleration after its transition accounts for the growth of $\log{\Delta E_{kin}}$ before the flare reconnection. In the text below, we refer to the phase after the transitions in the kinetic energy and acceleration as rapid exponential rise phase. Despite this transition, the order of magnitude of $\Delta E_{kin}$ before the flare reconnection ($<10^{31}$erg) is still significantly smaller than that after the eruption begins ($\approx 10^{31.5}$ erg), indicating that the main acceleration of the ejecta does not occur during the rapid exponential rise phase. The kinetic energy experiences a secondary transition near the flare reconnection onset, as $\Delta E_{kin}(t)$ transitions from the steady growth to a more rapid increase. The kinetic energy experiences its main increase after the flare reconnection, which is well after the rapid exponential rise phase for Cases 2 and 3. The analysis above shows that the rapid exponential rise phase does not correspond to the main acceleration, but rather indicates a trigger of the acceleration process.

The transition to a rapid exponential rise phase in Cases 2 and 3 is a strong indication of a self-amplifying process. However, several mechanisms may account for the transition to this phase. On one hand, when an ideal MHD instability, such as torus, occurs, the deviation of the system from the equilibrium grows exponentially with time \citep[][]{Torok2005,Kliem2006}. On the other hand, the rapid exponential rise phase may also be due to a reconnection-involved process. For example, when breakout reconnection occurs due to the radial motion of the MFR, the reconnection decreases the magnetic pressure and tension forces above the MFR, which in turn amplifies the acceleration. Therefore, the non-ideal reconnection may also generate a self-amplifying process, which leads to exponential growth of the motion, as in the classic resistive-kink instability \citep[e.g.,][]{Bateman1978}. The determination of the process that causes the rapid exponential rise phase will be discussed in the next section.

\section{CME Mechanisms \& The Effect of Background Field} \label{section_mechanism}

To understand the features and the differences among the three Cases presented above, we investigate the underlying CME mechanisms in this Section.

\subsection{Timing of Reconnection}

\begin{figure}[tp!]
\centering {\includegraphics[width=1.0\hsize]{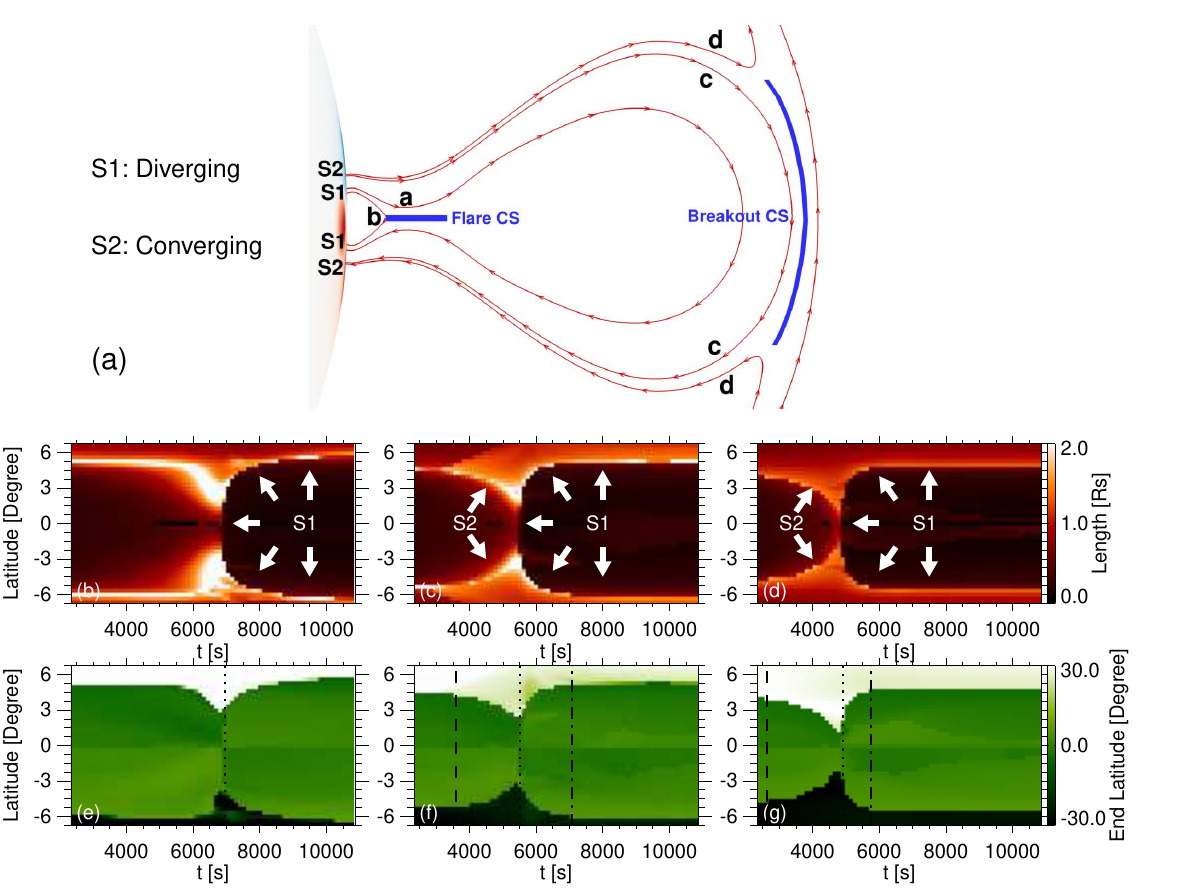}} 
\caption{(a): A two-dimensional sketch of the flare reconnection and breakout reconnection. (b)-(d): The time-latitude diagram of the field line length derived by tracing along the field line from a series of points lying on the intersection between the $y=0$ plane and the $1.0R_{s}$ sphere. (e)-(g): The time-latitude diagram of the end point latitude of the field line tracing. The three columns represent Cases 1 to 3, respectively. Dotted line: the onset time of the flare reconnection. Dotted-dashed line: the end time of the flare reconnection. Dashed line: the onset time of the breakout reconnection. } \label{fig_reconnection}
\end{figure}

Figure~\ref{fig_reconnection} (a) shows a two-dimensional sketch of the eruption. The timing of the reconnections, flare and breakout, can be determined by studying the behavior of several magnetic field lines anchored to the photosphere. In the flare reconnection, the magnetic field line labeled $a$ (strapping field line) transforms into the field line $b$ with a relatively short length (post-reconnection loop). At the instant of this transformation the reconnected field line defines a separatrix surface whose footpoints at the photosphere are labeled $S_{1}$. Similarly, the breakout reconnection transforms the field line $c$ to the two field lines labeled by $d$. The two points labeled by $S_{2}$ correspond to the separatrix associated with the breakout reconnection. During the two types of reconnections, the two points labeled by $S_{1}$ diverge due to the growth of the post-reconnection loop region, while the two points labeled by $S_{2}$ converge. Note that this sketch does not capture the full three-dimensional topology of the magnetic field, and it is shown only to illustrate the reconnection-associated features. We also note that this sketch only shows the breakout eruption before the separatrix $S_{2}$ contacts the separatrix $S_{1}$.

The features mentioned above can also be found in our three-dimensional simulations. Our method for analyzing reconnection is inspired by \citet{Dahlin2022b}. In each frame of the $120$ s data, we selected a series of uniformly distributed points lying on the intersection between the $y=0$ plane and the $1.01R_{s}$ sphere with a latitude step of $\Delta\lambda\approx0.33^{\circ}$. We traced along the magnetic field lines starting from these points until the field lines approached the boundary of the data box, and recorded the total length and the position of the end point of each field line. Figures~\ref{fig_reconnection} (b)-(g) present the time-latitude diagrams of the field line length and latitude of the end point derived from the field line tracing. In these diagrams, the boundaries with abrupt changes in value represent the footpoints of the separatrices. (Note, however that as discussed below, some of the boundaries are artifacts of our use of a finite sized box for performing the field line tracing.) The evolution of the separatrix footpoints reflects the occurrence of reconnection. We identify two types of boundaries that show the divergence of $S_{1}$ (flare reconnection) and the convergence of $S_{2}$ (breakout reconnection), respectively. The region inside the boundary labeled by $S_{1}$ corresponds to the post-reconnection loops (which are illustrated by field line $b$ in Figure~\ref{fig_reconnection} (a)). This region appears as an expanding region with a short field line length in each simulation. 

We determine the onset of the flare reconnection as the moment when this region appears, and label the moment with the vertical dotted line in Figures~\ref{fig_reconnection} (e)-(g). For Cases 1 to 3, the flare reconnection onset time is $t=6960$ [s], $t=5520$ [s], and $t=4920$ [s], respectively. We also notice that the post-reconnection loops region in Case 1 continues to expand to the end of the simulation. In contrast, the region in either Case 2 or Case 3 stops expanding during the simulation (labeled by the vertical dotted-dashed lines). This is consistent with the results in Figs.~\ref{fig_results} and ~\ref{fig_energy_curves} that in Case 1 the flare reconnection continues to a late phase in order to rebuild the closed helmet streamer to large heights; whereas in Cases 2 and 3 the flare reconnection need only continue until the coronal null is reestablished.  We then studied the early stage of the breakout reconnection in Cases 2 and 3 with the boundary labeled by $S_{2}$. We determined the onset time of the breakout reconnection as the moment when the convergence of $S_{2}$ begins (the vertical dashed lines), which yields $t=3600$ s and $t=2640$ s for Cases 2 and 3, respectively. 

One may notice that Figures~\ref{fig_reconnection} (b) and (e) exhibit a feature with long field line lengths and a jump of the end point latitude, which is similar to those labeled by $S_{2}$ at $\lambda\approx\pm 6^{\circ}$. As mentioned above, this feature is an artifact due to domain truncation rather than breakout reconnection. The feature indicates the boundary between the field lines that close within the box of the saved data and those that are open with regard to the box. From a topological perspective, the magnetic field of Case 1 prohibits the occurrence of the breakout reconnection and has no $S_2$ separatrix.

The comparison of the onset times derived above and the results in Figure~\ref{fig_speed} reveals that first, when the flare reconnection occurs, the MFR has already been accelerated and gained some measurable kinetic energy, and second, the breakout reconnection occurs earlier than the transition to the rapid exponential rise phase in both Cases 2 and 3. These suggest that the breakout reconnection is involved in the CME initiation, while the flare reconnection likely plays the major role in enhancing the eruption acceleration.  We note that our analysis of timing is affected by the resolution of the starting points for field line tracing. Since the latitudinal step of the staring points ($\approx0.33^{\circ}$) is larger than the minimum angular grid size in the simulation ($\approx0.11^{\circ}$), the diagrams in Figure~\ref{fig_reconnection} may miss finer features. Therefore, the actual onset time of the breakout reconnection can be earlier than the time we derived; however, this error does not affect the conclusion that the breakout reconnection occurs well before the transition to the rapid exponential rise phase.

\subsection{The Torus Instability}

\begin{figure}[tp!]
\centering {\includegraphics[width=1.0\hsize]{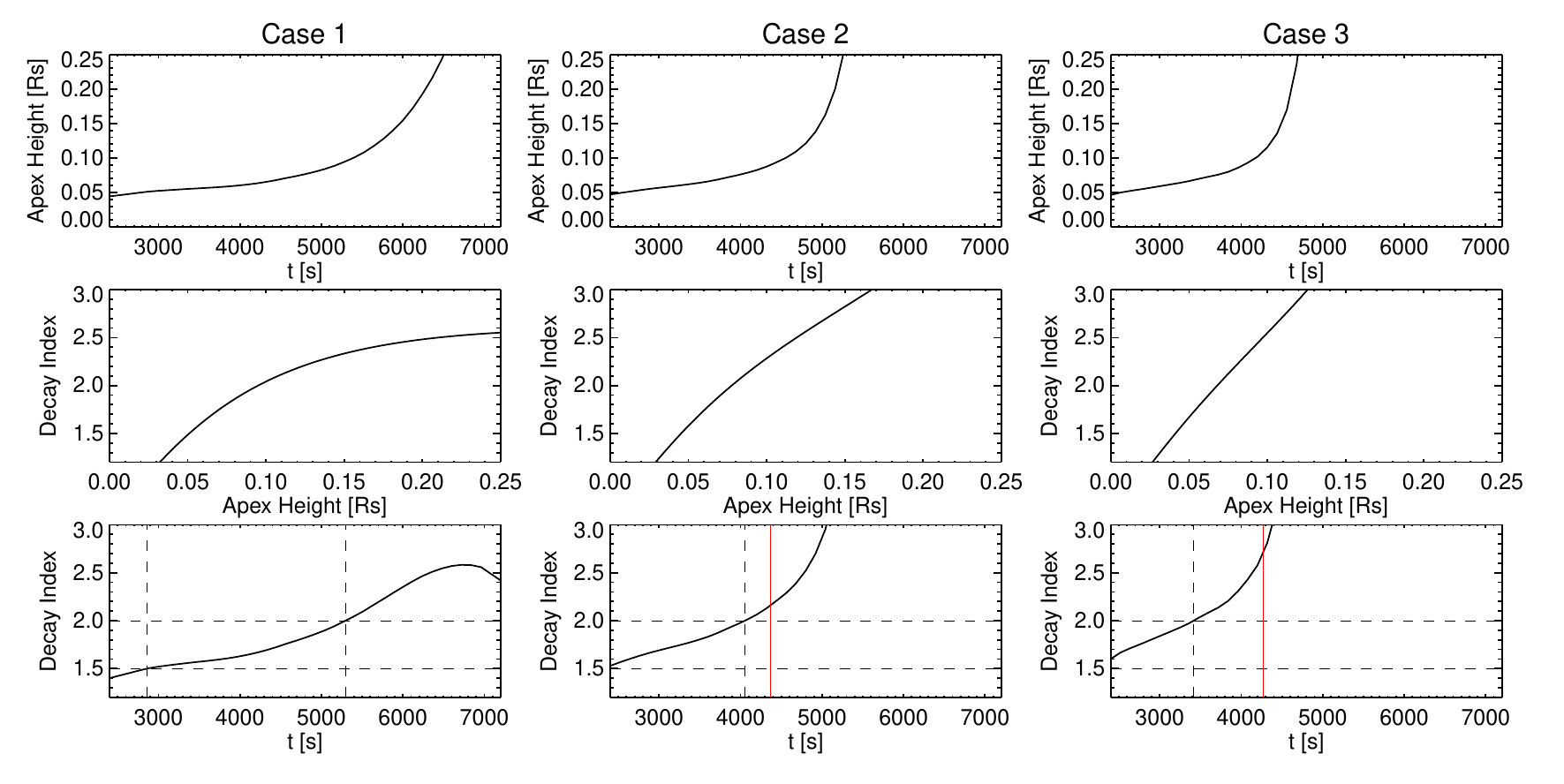}} 
\caption{Upper row: The MFR apex height ($H$) as a function of time. Middle row: The decay index as a function of height. Lower row: The decay index at the MFR apex. The three columns correspond to Cases 1 to 3, respectively. The red vertical lines in Cases 2 and 3 indicate the transition to the rapid exponential rise phase.} \label{fig_decay_index}
\end{figure}

The criterion for the torus instability is given by the decay index $n$:
\begin{equation}
    n=-\frac{\partial \ln{B_{ex}}}{\partial \ln{r_{c}}}. \label{eq_n}
\end{equation}
Here $r_{c}$ represents the distance from the MFR axis. \citet{Bateman1978} suggested that the torus instability occurs when $n$ is greater than the critical decay index $n_{cr}=1.5$. \citet{Kliem2006} derived a more accurate formula for the critical decay index $n_{cr}$, which yields $n_{cr}\approx1.5$. However, the calculation in these works focuses on the whole torus, while the coronal MFR is better described by a line-tied partial torus. Previous simulations of the MFR eruptions \citep[e.g., ][]{Fan2007,Fan2010,Aulanier2010,Torok2007} found that the $n_{cr}$ for a line-tied partial torus falls in the range $[1.5,2]$. We applied Equation~\ref{eq_n} to calculate the distribution of $n$ along the -$x$-axis in each simulation. Note that Equation~\ref{eq_n} is not applicable to the region above the null points in Cases 2 and 3. The bottom row of Figure~\ref{fig_decay_index} presents the decay index at the MFR apex as a function of time. We label the two times when $n=1.5$ and $n=2.0$, respectively, for Case 1. For Cases 2 and 3, the decay index is already $>1.5$ at $t=2400$ s, therefore we only label the time when $n=2.0$. For the three cases, the times when $n=2.0$ are $t\approx5290$s, $t\approx4050$s, and $t\approx3420$s, respectively. If the torus instability were present, then it should have occurred by this time. Each of these times is earlier than the corresponding flare reconnection onset. For either of Cases 2 and 3, the time when $n=2.0$ is also earlier than the transition to rapid exponential rise phase, which is labeled by the red solid line in Figure~\ref{fig_decay_index}.

A key point is the relation between the torus instability and the rapid exponential rise phase. For cases 2 and 3, the decay indices at the time when the system transitions to the rapid exponential rise phase are approximately $2.16$ and $2.72$, respectively. The two values are significantly higher than $n_{cr}$ (which lies in the range $[1.5,2]$), suggesting that the occurrence of the torus instability does not indicate the transition to the rapid exponential rise phase. The two values also differ by $\approx0.56$, which indicates that the transition to the rapid exponential rise phase does not correspond to a single value of the decay index. One may argue that the transition to the rapid exponential phase may be correlated with varying values of $n$ across Cases 1 to 3. With this assumption, due to the absence of the rapid exponential rise phase in Case 1, the required decay index in Case 1 for the rapid exponential rise phase would be greater than the maximum value of the decay index in Case 1, i.e., $2.58$. Considering that the decay indices at the onset of the rapid exponential rise phase in Cases 2 and 3 are $2.16$ and $2.72$ respectively, one would obtain a non-monotonic trend of the required decay index from Cases 1 to 3, which is inconsistent with the monotonic variation of the background field strength from Cases 1 to 3. Therefore, we deduce that the transition to rapid exponential rise phase does not exhibit a correlation with the criterion for torus instability.

\subsection{Discussion}

\begin{figure}[tp!]
\centering {\includegraphics[width=1.0\hsize]{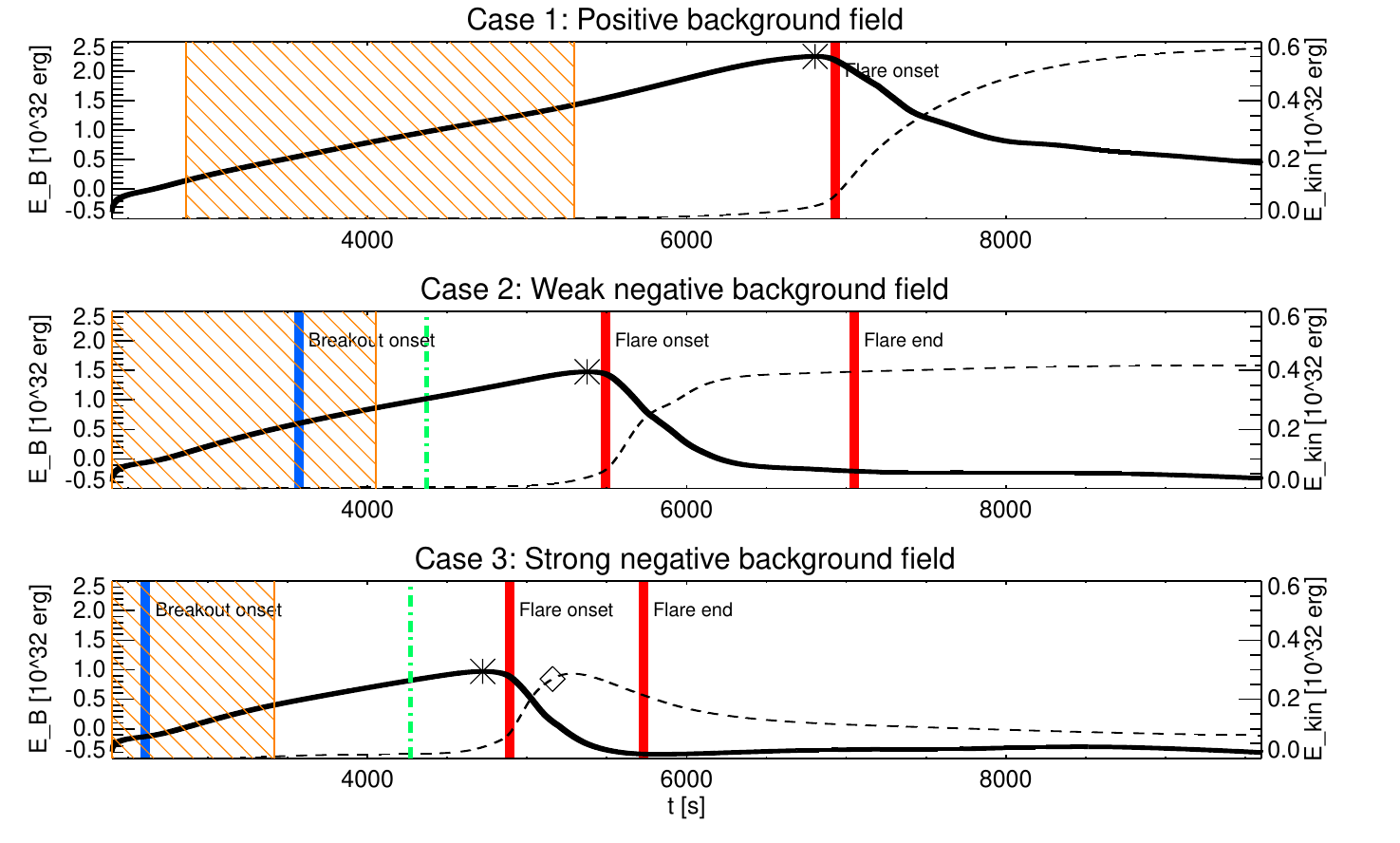}} 
\caption{The timing of processes of interest and the energy curves for three cases. In each panel, the black solid curve and the black dashed curve indicate $E_{B}(t)$ and $E_{kin}$, respectively. The scales for $E_{B}(t)$ and $E_{kin}$ are given by the left and right vertical axes, respectively. The red vertical lines label the onset and end of the flare reconnection. For each of Cases 2 and 3, the blue vertical lines label the onset of the breakout reconnection. The orange region corresponds to the interval $1.5<n<2.0$. The green dotted-dashed vertical lines label the transition to the rapid exponential rise phase in Cases 2 and 3..} \label{fig_timing}
\end{figure}

Figure~\ref{fig_timing} summarizes the timing of the rapid exponential rise phase, reconnections, and the decay index variation, along with the $E_{B}(t)$ and $E_{kin}$ curves of each case. We label the transition to rapid exponential rise phase with the green dotted-dashed lines for Cases 2 and 3. The breakout and flare reconnections are labeled by the blue and red vertical lines, respectively. The orange spaced lines indicate the interval of $1.5<n<2.0$ for each case. Our previous analysis suggest the following conclusions. (a) The onset of flare reconnection occurs after but very near the onset the MFR acceleration. (b) The breakout reconnection occurs during the rapid exponential rise phase of either Cases 2 or 3. Both breakout reconnection and the rapid exponential rise phase are absent in Case 1. (c) The torus instability criterion is satisfied in each case at an early stage of the eruption. However, the transition to the rapid exponential rise phase does not exhibit a clear correlation with the criterion for the torus instability. Based on these conclusions, we suggest that the breakout reconnection is most likely the process responsible for the rapid exponential rise phase in Cases 2 and 3. The exponential growth rate in Case 3 ($\alpha=0.0029$ s$^{-1}$) is higher than that in Case 2 ($\alpha=0.0018$ s$^{-1}$), which is consistent with our finding in Figure~\ref{fig_reconnection} that the reconnection in Case 3 is earlier and more rapid than that in Case 2. Although the ideal torus instability is also capable of causing an exponential growth of the motion, the comparison of Case 1 with the other cases indicates that the contribution from the torus instability is not clear in our simulations. In Case 1, the torus instability may be the main physical process that leads to the growth of the acceleration and kinetic energy; however, due to the absence of the breakout reconnection, the growth rate in Case 1 is significantly lower than those in the rapid exponential rise phases in Cases 2 and 3. Based on the comparisons above, we find that the decrease in the background field leads to a faster growth of the acceleration of MFR in the early stage of the eruption.

Another important aspect is the energy release phase of the eruptions. Figure~\ref{fig_timing} suggests that the flare reconnection and the release of the magnetic energy are relatively late compared to the initial acceleration. The flare reconnection onset marks the onset of a rapid conversion of magnetic energy to kinetic energy. Before the flare reconnection onset, the kinetic energy $E_{kin}(t)$ increases slowly due to the early acceleration of the MFR. The magnetic energy $E_{B}$ does not decrease significantly before the onset of flare reconnection, and it continues to increase due to the STITCH driving. After the flare reconnection begins, magnetic energy rapidly converts to kinetic energy.  In Case 1, the flare reconnection does not terminate at the end of the simulation, while in Cases 2 and 3, the flare reconnection ends at $t\approx 7000$ s and $t\approx 5700$ s, respectively. The release of $E_{B}$ in Case 1 is maintained to the end of the simulation, whereas in Cases 2 and 3, the release of $E_{B}$ is weak after the flare reconnection ends. The features above reveal a strong correlation between the occurrence of the flare reconnection and the release of the magnetic free energy.

The qualitative differences in the flare reconnection and energy release between Case 1 and the other two cases can be explained by differences in the external magnetic field topology. In case 1, the flare reconnection produces a helmet-streamer as the post-eruptive magnetic configuration. The CS of the helmet-streamer connects the CME MFR and the post-reconnection loops. In this way, the flare reconnection is maintained as the eruption proceeds. In Cases 2 and 3, the flare CS evolves into a new coronal null line. When the flare CS approaches the position of the new null line, the CS stops the radial elongation and is decoupled from the MFR. Consequently, the flare reconnection is terminated. The analysis above suggests that the background field affects not only the mechanisms in the early stage of the eruption, but also the energy-release process on a longer temporal scale.

Besides comparing our simulations, it is also interesting to compare our results with those in \citet{Karpen2012}. In their simulation of an SMA eruption, they found that the SMA first experiences an early acceleration, followed by a temporary stall until the flare reconnection produces the first plasma jet (a plasmoid with an O-type point) that propagates outward. The plasma jet drives fast flare reconnection, which they suggested is the primary physical process that initiates the rapid acceleration of the CME. One reason for the discrepancy between their results and our results may be the different choices of pre-eruptive configuration (SMA or MFR). In our simulations, the inserted MFR has a topology similar to that of the plasma jet, indicating that the erupting MFR itself can drive fast flare reconnection. Therefore, as long as the MFR motion initiates flare reconnection, the self-driven fast flare reconnection begins, leading to rapid energy conversion.

\section{Summary and Conclusions} \label{section_summary}

In this work, we performed three AWSoM-R simulations of the MFR eruptions with different background magnetic fields. We used the mTD and STITCH models to simulate the pre-eruptive and trigger phases of the eruptions. We produced three MFR eruptions with different magnetic field evolutions. In Case 1, the eruption only involves the flare reconnection. In Case 2, both flare reconnection and breakout reconnection are involved. The magnetic field evolution in Case 3 is similar to that in Case 2 during the early stage of the eruption. However, the breakout reconnection later erodes the MFR in Case 3, leading to a failed eruption. We analyzed the evolution of magnetic and kinetic energy in each simulation and studied the dynamics of the MFR. To understand the features and the differences among simulations, we then investigated the timing of reconnections and the criterion for the torus instability.

In general, we found that the background magnetic field (characterized by $B_{BG,r}^{N}$) affects the eruption in such a way that as $B_{BG,r}^{N}$ decreases, the acceleration growth rate increases and the energy release becomes earlier. The free energy required to initiate the eruption shows a positive correlation with the background field. One of our key findings is that in Cases 2 and 3, the kinetic energy and acceleration both exhibit a transition from a relatively slow growth to a rapid exponential rise phase. The breakout reconnection likely accounts for the rapid exponential rise phase. Our discussion showed that the rapid exponential rise phase does not exhibit a clear correlation with the criterion for torus instability. Another interesting result is that the background magnetic field affects the onset and duration of the flare reconnection. The flare reconnection onset becomes earlier as $B_{BG,r}^{N}$ decreases. A positive value of $B_{BG,r}^{N}$ allows for a long-existing flare reconnection, while a negative value of $B_{BG,r}^{N}$ leads to the termination of the flare reconnection during the eruption. We also notice that the simulation results for Case 1 exhibit significant differences from those of Cases 2 and 3. First, the rapid exponential rise phase seen in Cases 2 and 3 is absent in Case 1. Second, the flare reconnection in Case 1 is maintained for a long duration, while the flare CS in Cases 2 and 3 becomes a new null-point after a short duration of reconnection. We point out that both differences are due to the topological difference between Case 1 and the other two cases.

An important feature that is missing in the work above is rigorous thermodynamics in the reconnection CS. In our simulation and in most previous simulations \citep[e.g.,][]{Karpen2012,Torok2018}, energy conservation is not satisfied due to numerical diffusion, especially near the reconnection CS. Consequently, the released magnetic free energy is only partially converted to kinetic energy, and the part that should have been converted to thermal energy is lost. This could well lead to an underestimation of plasma $\beta$ in the CS, which would significantly affect the reconnection dynamics. Our future work will address this issue and simulate eruptions with more realistic reconnection.

\begin{acknowledgments}
We acknowledge support from the NASA SCEPTER LWS Strategic Capability Program, the NASA CLEAR Space Weather Center Program, the NSF ANSWERS grant GEO-2149771, and NSF grant 2229337 to the University of Michigan. N. Sachdeva also acknowledges support from the NASA LWS grant 80NSSC24K1104. XYL thanks Dr. Bernhard Kliem and Dr. Benjamin Lynch for valuable discussions. 
\end{acknowledgments}

\bibliographystyle{aasjournalv7}
\bibliography{references}

\end{document}